\documentclass[preprint,endfloats*,else,gca]{revtex4-2}
\usepackage{natbib}
\usepackage[dvips]{graphicx}
\usepackage{amssymb}
\usepackage{threeparttable}
\usepackage{longtable}
\usepackage{amsmath}
\usepackage{amsfonts}
\usepackage{verbatim}
\usepackage{wasysym}
\usepackage{bm}
\usepackage{dcolumn}% Align table columns on decimal point
\usepackage[usenames]{color} % \textcolor{Red}{text} makes text red
\makeatletter
\def\@dotsep{4.5}
\renewcommand\subsection{\@startsection
  {subsection}{2}{0mm}%name, level, indent
  {-\baselineskip}%             beforeskip
  {0.5\baselineskip}%            afterskip
  {\normalfont\normalsize\bfseries}}% style
\renewcommand\subsubsection{\@startsection
  {subsubsection}{3}{0mm}%name, level, indent
  {-\baselineskip}%             beforeskip
  {0.2\baselineskip}%            afterskip
  {\normalfont\normalsize\itshape}}% style

\bibpunct{}{}{,}{a}{}{\textsuperscript{,}}% enleve la numerotation de la biblio
\makeatother
\begin{document}
\renewcommand{\thetable}{\arabic{table}}
\renewcommand{\thesection}{\arabic{section}}
\renewcommand{\thesubsection}{\arabic{section}.\arabic{subsection}}
\renewcommand{\thesubsubsection}{\arabic{section}.\arabic{subsection}.\arabic{subsubsection}}
\renewcommand{\theparagraph}{\arabic{section}.\arabic{subsection}.\arabic{subsubsection}.\arabic{paragraph}}
\newcommand{\citemt}[1]{\citeauthor{#1}~(\citeyear{#1})}
\newcommand{\citemp}[1]{\citeauthor{#1},~\citeyear{#1}}
\title{From Oxygen to Silicon equilibrium isotopic fractionation:
 assessments from first-principles density-functional theory.}
\author{Merlin M\'eheut$^1$, Michele Lazzeri$^1$, 
Etienne Balan$^{1,2}$, and Francesco Mauri$^1$}
\affiliation{
$^1$ IMPMC, Universit\'e Paris VI et VII, CNRS, IPGP, 4
Place Jussieu, 75252, Paris cedex 05, France \\
$^2$ IRD -UMR CEREGE , Europole M\'editerran\'een de l'Arbois, 
BP 80 , 13545 Aix en Provence cedex, France}
%\date{\today, Revised Version}
%\date{\today}
\date{November 16, 2007}

\begin{abstract}
	Isotopic fractionation factors for oxygen and silicon
in selected silicates (quartz, enstatite, forsterite, lizardite, kaolinite)
 have been calculated using first-principles methods. 
Good agreement between theory and experiment is 
obtained in the case of oxygen. In the case 
of silicon, agreement and differences with existing estimates of equilibrium 
fractionation factors are discussed. The relationships 
between silicon fractionation factors
and oxygen fractionation factors, silicate polymerization degree and chemical composition is studied. 
The previously stated relationship with the polymerization degree of the mineral is not confirmed. 
Nevertheless, our calculation suggests that silicon fractionation depends on the cationic content of the mineral, in a 
way similar to oxygen. Oxygen fractionation properties could thus 
provide some insights on silicon fractionation behavior. 
\end{abstract}

\maketitle

\pagebreak  %% a mettre si preprint, a enlever sinon

\bibliographystyle{gca}

\section{Introduction}

	Silicon isotopic studies have been carried out for over half a century (\citemp{Reynold1953}; \citemp{Epstein1970};\citemp{Clayton1978}; 
\citemp{Douthitt1982}; \citemp{Ding1996}; \citemp{DeLaRocha2000}; 
\citemp{Georg2007}). A large number of 
terrestrial and extraterrestrial samples were investigated and the general picture of silicon isotope distribution in nature 
was outlined (\citemp{Douthitt1982}; \citemp{Ding1996}). 

The largest variations of silicon isotopes in terrestrial samples 
have been found in surface environments and under low temperature conditions 
(\citemp{Douthitt1982}; \citemp{Ding1996}; \citemp{DeLaRocha2000}). 
Isotopic fractionation is therefore an important tool for understanding the 
continental cycle of silicon (\citemp{DeLaRocha1997},\citeyear{DeLaRocha2000};
\citemp{Basile-Doelsch2005}). In those surface processes,
the exchange to be considered is between the mineral (quartz,clays,opal) and silicic acid in solution.
More specifically, silicon fractionation appearing during the precipitation
of clays successive to the weathering of crustal rocks 
 is thought to play a role in the isotopic composition of sea and river waters 
(\citemp{DeLaRocha2000}; \citemp{Ding2004}; \citemp{Georg2007}). 
It is however difficult to conclude about the nature 
(kinetic or equilibrium) of the fractionation process
(\citemp{DeLaRocha1997}). 

	On the other hand, igneous rocks show limited but measurable (1.1$\permil$, close to experimental accuracy)
$\delta^{30}Si$ variations, which present systematic trends that led to postulate equilibrium fractionation 
processes (\citemp{Douthitt1982}). Several trends have been proposed: 
following \citemt{Douthitt1982}, coexisting minerals in igneous
rocks exhibit small, systematic silicon isotopic fractionations 
that are roughly 1/3 the magnitude of concomitant oxygen isotopic 
fractionations at 1150$^\circ$C, and
$\delta^{30}Si$ shows a positive correlation with silicon content. 
Following \citemt{Ding1996}, the ${}^{30}Si$ content
may increase as a function of the degree of polymerization of the 
SiO$_4$ tetrahedrons in the structure, in the same way as that of ${}^{18}O$. 
In another words,
the fractionation should be positive between a silicate and another 
of lower polymerization degree.
This last assumption is supported by theoretical considerations, 
such as the early calculations of
\citemt{Grant1954} on the inosilicate-tectosilicate system. 
These calculations are based on the assumption
that silicon fractionation is mainly determined by the 
Si-O stretching frequencies, which increase
with the degree of polymerization and the number of oxygen bonds 
between Si-O tetrahedrons.
Nevertheless, if qualitatively satisfactory, the study of Grant greatly 
overestimates the fractionation
factor with respect to observations.
That is why, as pointed out by \citemt{Ding1996}, 
obtaining a reliable theoretical prediction relating
the structure of silicates and their fractionation properties 
with respect to silicon is an
important aspect of silicon isotopic geochemical studies. 
Moreover, analytical precision has been yet insufficient to assess any measurable
temperature dependence of fractionation factors, rendering desirable the use of theoretical methods
to predict this dependence. Recent analytical development in mass-spectrometry will open numerous
possibilities to learn, beyond the study of the range of natural variations of these stable isotopes,
what they tell us about geological processes.
The aim of this study is thus to evaluate the
possible relationship between the silicate structure and their silicon content, oxygen fractionation
properties and silicon fractionation properties, at a theoretical level.
Within the framework of first-principles density functional theory, we have recently developed a methodology to
predict equilibrium fractionation factors as a function of temperature (\citemp{Meheut2007}).
In the present study, we calculate the fractionation properties of oxygen and silicon for several minerals 
displaying various degrees of polymerization of silicate units. Polymerization degree is noted Q$^\textrm{n}$,
where n is the number of bridging oxygens for one SiO$_4$ unit (\citemp{Engelhardt1975}). 
Structures studied here are  quartz (tectosilicate, Q$^\textrm{4}$ ), 
lizardite (phyllosilicate, Q$^\textrm{3}$), enstatite (inosilicate, Q$^\textrm{2}$), 
and forsterite (nesosilicate, Q$^\textrm{0}$). 
From the assumption made by \citemt{Douthitt1982} that quartz-dissolved silicon equilibrium fractionation 
factor is 0$\permil$ at temperatures as low as 500$^\circ$C,  we also use our previous calculation on kaolinite 
(\citemp{Meheut2007}) to assess the Si fractionation between clay minerals and solute Si species.

\section{Methods}

The methods used here have been exposed in details in \citemt{Meheut2007}.

\subsection{The Isotopic Fractionation Factor $\alpha$.}

The $\beta$-factor $\beta(a,Y)$ is the isotopic fractionation factor of element Y 
between the phase $a$ and a perfect gas of $Y$ atoms, having the natural mean isotopic concentration.
The isotopic fractionation factor $\alpha(a,b,Y)$ relative to an atom $Y$, between two phases $a$ and $b$ 
can be written as the ratio of the $\beta$-factors relative to this atom and to each phase separately
(\citemp{Richet1977}). 
The results will be discussed in terms of the logarithmic $\beta$-factors ($\ln \beta$)
and logarithmic fractionation factors ($\ln\alpha$) expressed
in parts per thousand ($\permil$).

\subsection{DFT calculations.} 
	
	The $\beta$-factors are computed using Eqs. (6) and (8) of \citemp{Meheut2007}
 (the difference between the results of the two equations is less than 0.1$\permil$ in all systems considered here) 
from the harmonic partition function, which, in turn, is obtained from Eq. (16) 
of \citemp{Meheut2007}, as a function of the phonon frequencies.
	
	The phonon frequencies are computed using first-principles methods based on density
functional theory (DFT) (\citemp{Hohenberg1964}; \citemp{Kohn1965}).
We use the generalized-gradient approximation
to the exchange-correlation functional of Perdew, Burke and
Ernzerhof (PBE) (\citemp{Perdew1996}).  The ionic cores are described by
norm-conserving pseudopotentials (\citemp{Troullier1991}) in the
Kleinman-Bylander form (\citemp{Kleinman1982}).
The description of the pseudopotentials is given in the electronic
annexes (Table EA-1). For 
quartz and kaolinite, we used our former calculations described in \citemt{Meheut2007}.
For lizardite, forsterite and clinoenstatite, the electronic wave-functions are expanded
in plane-waves up to an energy cut-off $\epsilon_{cut}=$150~Ry and
the charge density cut-off is $4\epsilon_{cut}$. The electronic integration
is performed by sampling the Brillouin zone with a 2$\times$2$\times$2
k-point grid for lizardite and forsterite, according to the Monkhorst-Pack scheme
(\citemp{Monkhorst1976}). For clinoenstatite, the Brillouin zone sampling is
restricted to the Baldereschi point (1/4, 1/2, 1/4) (\citemp{Baldereschi1973}).  
% expliquer? 

	In all the cases, atomic positions are obtained after
relaxation at zero pressure until the residual forces
are less than $10^{-3}$ Ry/ \AA. Neither the cell parameters nor
the symmetry are constrained during the relaxation.

	The vibrational properties are calculated using the linear
response theory (\citemp{Baroni2001}), using the PWSCF package (Baroni et al.,
http://www.pwscf.org). The phonon frequencies 
are obtained following the standard procedure. First, we compute the dynamical
matrices exactly (within DFT) on a regular grid of q-vectors.
These matrices are then used to determine the interatomic force-constants.
Subsequently, the dynamical matrix can be determined
in any point of the Brillouin zone by discrete Fourier interpolation of the
interatomic force-constants.
Long-range effects are taken into account after computing the Born
effective-charges and the dielectric constant following ~\citemt{Baroni2001}.
The treatment is exact for a sufficiently large q-vector grid.
The dynamical matrices are computed exactly
(within DFT) on a 2$\times$2$\times$2 q-point grid for lizardite,
on a 1$\times$1$\times$1 grid for forsterite and clinoenstatite (i.e. the 
matrix of force constants has been generated with the dynamical matrix at gamma). 
	Then, the vibrational
partition function (Eq.(16) of \citemp{Meheut2007}) is obtained performing the product
 on a n$\times$n$\times$n grid, with $n$=5 for forsterite, enstatite and lizardite.

\section{Results and discussion}

\subsection{Relaxed Structures.}

\subsubsection{Structures formerly studied.}

Agreement between the relaxed structures of kaolinite (Table EA-2 of \citemp{Meheut2007}), lizardite 
(Table \ref{tab:strucliz} and Table A-2 of additional material) , quartz (Table EA-1 of \citemp{Meheut2007}), 
and experiment are discussed  in \citemt{Balan2001a}, \citemt{Balan2002a} and \citemt{Meheut2007}.

        \begin{table}[!hbt]
        \caption{Cell parameters of lizardite}
        \label{tab:strucliz}
         \begin{threeparttable}[b]
          \begin{tabular}{rcl}
        \hline
        & This work (PBE)   & Expt.\tnote{$\dagger$}  \\ \hline
        a (\AA)            &  5.3743          &   5.3267   \\
        c (\AA)            &  7.4456          &   7.2539   \\
        \hline
          \end{tabular}
          \begin{tablenotes}
            \item [$\dagger$] \citemt{Gregorkiewitz1996}
           \end{tablenotes}
         \end{threeparttable}
        \end{table}

\subsubsection{Forsterite.}

The structure of forsterite was relaxed at the PBE level. 
The cell parameters of the relaxed structure are shown Table \ref{tab:strucforst}.
They show good agreement with experiment and consistency with previous computations (see e.g. \citemp{Noel2006}).
Internal coordinates of the relaxed structure are given Table A-3 in additional material.
%% voire refs sur enstatite relaxee et discuter l'accord theorie-experience
%% dans les cas ou ce n'a pas ete deja fait

       \begin{table}[!hbt]
        \caption{Cell parameters of forsterite}
        \label{tab:strucforst}
         \begin{threeparttable}[b]
          \begin{tabular}{rcl}
        \hline
        & This work (PBE)   & Expt.\tnote{$\dagger$}  \\ \hline
        a (\AA)            &  4.8032          &    4.7530    \\
        b (\AA)            &  10.3321          &   10.1900    \\
        c (\AA)            &  6.0448           &    5.978    \\
        \hline
          \end{tabular}
          \begin{tablenotes}
            \item [$\dagger$] \citemt{Fujino1981}
           \end{tablenotes}
         \end{threeparttable}
        \end{table}

\subsubsection{Enstatite.}

Enstatite $Mg_2Si_2O_6$ has four polymorphs: protoenstatite, orthoenstatite, and low 
and high clinoenstatite, the structures of which being closely related. The basic building blocks are 
single tetrahedral silicate chains and double [MgO$_6$] octahedral bands running parallel to the c axis. 
Structurally, the polymorphs are distinguished by the various stacking sequences depending on the 
orientation of the [MgO$_6$]
octahedra with respect to the silicate chains. Because of their close structural similarities, 
the lattice-dynamical and thermodynamic properties of the enstatite polymorphs are expected 
to be very similar. Ortho-enstatite is the most common 
form at ambient pressure, and the mineral on which we have the most vibrational and isotopic properties. Unfortunately, the 
size of the unit cell (80 atoms) of this material drastically limits the possibilities of calculation. Consequently, 
we have chosen in this study to calculate the low-P clinoenstatite structure (40 atoms per unit cell) as a model for the 
thermodynamic properties of orthoenstatite. This structure is present in Nature, but is expected to be a metastable 
form coming from the quenching of High-P clinoenstatite. Cell parameters of our relaxed structure 
are compared with experimental data
in Table \ref{tab:strucenst}. Internal coordinates of the relaxed structure 
are given Table A-4 in additional material.

        \begin{table}[!hbt]
        \caption{Cell parameters of clinoenstatite}
        \label{tab:strucenst}
         \begin{threeparttable}[b]
          \begin{tabular}{rcl}
        \hline
        & This work (PBE)   & Expt.\tnote{$\dagger$}  \\ \hline
        a (\AA)            &  5.2394  & 5.188   \\
        b (\AA)            &  9.7450  & 9.620 \\
        c (\AA)            &  8.9486  & 8.825 \\
        $\gamma$ ($\,^{\circ}$)& 108.67 & 108.20    \\
        \hline
          \end{tabular}
          \begin{tablenotes}
            \item [$\dagger$] \citemt{Morimoto1960} 
           \end{tablenotes}
         \end{threeparttable}
        \end{table}

\subsection{Vibrational properties}

Agreement between the measured and computed vibrational properties of kaolinite,
lizardite and quartz are discussed in \citemt{Balan2001a}, 
\citemt{Balan2002a} and \citemt{Meheut2007}.

The calculated vibrational frequencies of forsterite are reported in Tables \ref{tab:vibforstIR} and 
\ref{tab:vibforstRaman}. The frequencies 
calculated by \citemt{Noel2006} showed good agreement with experimental frequencies, with a slight overestimate typical of 
B3LYP-type functionals. Our calculation underestimates experimental frequencies by around 5\% as usually observed 
in calculations using the PBE approximation. The experimental results are from \citemt{Reynard1991} 
and \citemt{Chopelas1991}, with the attribution of the 
modes done as exposed in \citemt{Noel2006}.

\begin{figure}[!hbp]
        {\includegraphics[width=120mm]{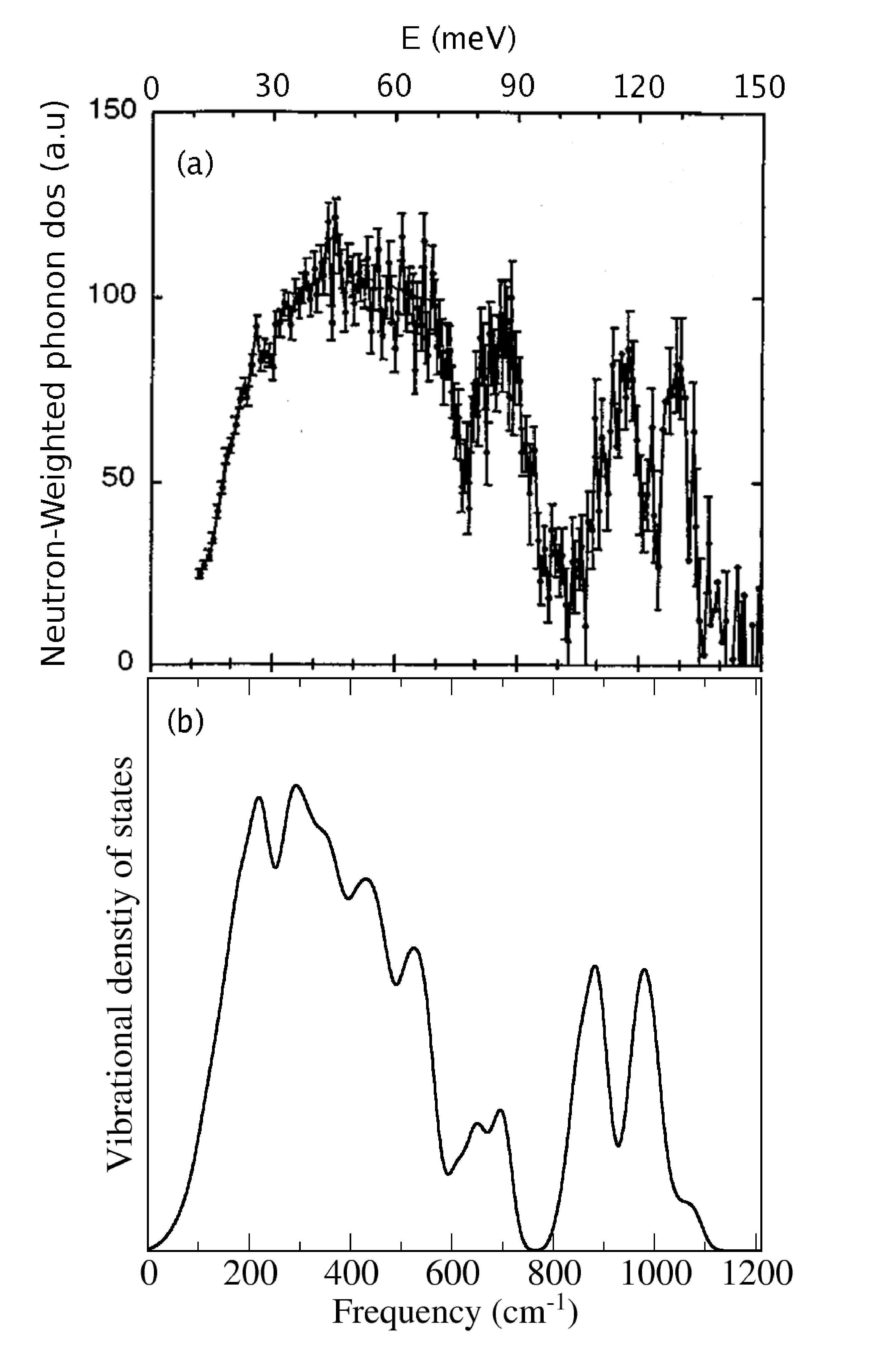}}
        \caption{Comparison between the calculated vibrational density of states of clinoenstatite (b) 
	and the IINS spectrum
 	of orthoenstatite from \citemt{Choudhury1998} (a). The x-axes on the two panels are the same.}
        \label{fig:enstatite_vdos}
        \end{figure}

The calculated vibrational density of states of clinoenstatite is compared in Fig.~\ref{fig:enstatite_vdos} with the
vibrational density of states of orthoenstatite estimated from 
inelastic incoherent neutron scattering measurements by \citemt{Choudhury1998} on orthoenstatite. 
Our calculated curve has been broadened by interpolation with a Gaussian function of FWHM 4.6 meV, to be compared with the 
assessed instrumental resolution, estimated to be 4-12 meV depending on the scattered-neutron energy. 
A detailed discussion of the analysis of the inelastic-neutron-scattering data in the incoherent approximation 
for comparison with the phonon density of states is given in \citemt{Rao1988} (see also \citemp{Fultz2004}).
The characteristics of the experimental 
measurements are well reproduced by our calculation. The most obvious difference is a frequency shift of -5\% between our 
calculation and experiment, consistently with former calculations.  

          \begin{longtable}{rcccccl}
	 \caption{Comparison of calculated and measured IR active TO and LO modes in forsterite\label{tab:vibforstIR}}\\
        \hline
&\multicolumn{2}{c}{Our Work (PBE)} & \multicolumn{2}{c}{ B3LYP$^\dagger$} & \multicolumn{2}{c}{ Exp. $^\ddagger$} \\
\hline 
 &  TO & LO & TO & LO & TO & LO \\
\endfirsthead
\caption{(continued)}\\
&\multicolumn{2}{c}{Our Work (PBE)} & \multicolumn{2}{c}{ B3LYP$^\dagger$} & \multicolumn{2}{c}{ Exp. $^\ddagger$} \\
\hline
 &  TO & LO & TO & LO & TO & LO \\
\endhead
\hline
\multicolumn{7}{l}{$\dagger$ \citemt{Noel2006} }\\
\multicolumn{7}{l}{$\ddagger$ \citemt{Reynard1991}}\\
\endfoot
B$_{1u}$ & & & & & & \\
& 193  &          &  207  &  207 &     & \\ 
& 261  &   261 &  278  &  279 & 274 & 277 \\ 
& 273  &   298 &  290  &  313 & 282 & 307 \\  
& 284  &   284 &  313  &  320 &     & \\ 
& 387  &   390 &  420  &  426 & 403 & 410 \\ 
& 398  &   437 &  428  &  461 & 411 & 451 \\ 
& 457  &   463 &  490  &  499 & 472 & 482 \\ 
& 483  &   553 &  514  &  592 & 501 & 580\\ 
& 830  &   947 &  874  &  1005& 869 & 1008 \\ 
B$_{2u}$ & & & & & & \\
& 135  &   136 &  143  &   144 & 140 & 143 \\  
& 261  &   262 &  277  &   277 & 275 & 278 \\ 
& 272  &   293 &  292  &   312 & 283 & 290 \\ 
& 329  &   357 &  350  &   387 & 341 & 371 \\ 
& 376  &   389 &  403  &   417 & 390 & 406 \\ 
& 396  &   423 &  432  &   453 & 415 & 439 \\ 
& 437  &   469 &  465  &   495 & 456 & 484 \\ 
& 485  &   487 &  517  &   520 & 503 & 511 \\ 
& 504  &   545 &  535  &   588 & 526 & 572 \\ 
& 599  &   599 &  638  &   638 &     & \\ 
& 791  &   796 &  835  &   843 & 830 & 841 \\ 
& 829  &   915 &  870  &   966 & 867 & 970 \\ 
& 934  &   943 &  989  &   999 & 987 & 1001 \\
B$_{3u}$ & & & & & &          \\ 
& 192  &   193 &  206  &   207 & 199 & 201 \\  
& 260  &   261 &  275  &   276 & 274 & 276\\  
& 277  &   282 &  294  &   300 &     &\\  
& 304  &   305 &  322  &   323 & 316 & 320 \\  
& 359  &   366 &  388  &   398 & 374 & 384 \\  
& 382  &   448 &  412  &   473 & 397 & 463 \\  
& 452  &   451 &  476  &   482 &     &\\  
& 482  &   507 &  513  &   539 & 498 & 544 \\  
& 507  &   529 &  540  &   563 &     &\\  
& 577  &   616 &  614  &   660 & 601 & 650 \\  
& 792  &   793 &  838  &   838 & 833 & 839 \\  
& 911  &   918 &  962  &   971 & 952 & 961 \\  
& 927  &  1023 &  982  &  1086 & 971 & 1089\\  
 \hline                        
          \end{longtable}             

          \begin{longtable}{rccl}
\caption{Comparison of calculated and measured frequencies of the Raman-active modes in forsterite\label{tab:vibforstRaman}}\\
        \hline
& Our Work (PBE) &  B3LYP $^\dagger$ &  Exp. $^\ddagger$ \\
\hline
\endfirsthead
\caption{(continued)}\\
 \hline
& Our Work (PBE) &  B3LYP $^\dagger$ &  Exp. $^\ddagger$ \\
\hline
\endhead
\hline
\multicolumn{4}{l}{$\dagger$ \citemt{Noel2006} }\\
\multicolumn{4}{l}{$\ddagger$ \citemt{Chopelas1991}}\\
\endfoot
A$_g$ & & &       \\
&  176 &   188  & 183  \\   
&  219 &   234  & 226  \\   
&  289 &   307  & 304  \\   
&  310 &   329  & 329  \\   
&  322 &   345  & 332  \\   
&  400 &   425  & 422  \\   
&  520 &   560  & 545  \\   
&  579 &   618  & 608  \\   
&  781 &   819  & 824  \\   
&  812 &   856  & 856  \\ 
&  914 &   967  & 965   \\
B$_{1g}$ & & &     \\
&  210 &   225 & 220 \\
&  242 &   260 & 274\\
&  301 &   317 & 318\\
&  335 &   367 & 351\\
&  371 &   391 & 383\\
&  415 &   442 & 434\\
&  552 &   596 & 582\\
&  600 &   645 & 632\\
&  793 &   835 & 838\\
&  820 &   866 & 866\\ 
&  925 &   979 & 975\\
B$_{2g}$ & & &           \\
&  170 &   183 &  175 \\
&  230 &   253 &  242\\
&  305 &   324 &  323\\
&  347 &   373 &  365\\
&  419 &   451 &  439\\
&  558 &   608 &  586\\
&  837 &   883 &  881 \\
B$_{3g}$ & & &          \\                 
&  176 &    190 & \\
&  275 &    303 &  286 \\
&  302 &    322 &  315 \\
&  356 &    381 &  374 \\
&  393 &    421 &  410 \\
&  564 &    609 &  592 \\
&  875 &    927 &  920 \\
	\hline                       
          \end{longtable}

\subsection{Isotopic Fractionation}

 Logarithmic $\beta$-factors have been computed every 0.5 $\frac{10^6}{T^2}$ (in K$^{-2}$).
Those points have been fitted with third order polynoms over different
domains of temperature. The domain of interest here is 0-1500$^{\circ}$C for the
 mineral phases.
The temperature domains have been cut in half for taking into account the non-linear behavior 
of the curves: 0-400$^{\circ}$C and 400-1500$^{\circ}$C.

\begin{table*}
\begin{center}
\caption{Fits of 1000$\ln \alpha$ based on the function
$a + bx + cx^2 + dx^3$, with $x = 10^6/T^2$.}
\label{tab:fits}
\begin{threeparttable}
\begin{tabular}{rcccccl}
\hline \hline
element         & phases        & T ($^\circ$C)     & $a$     & $b$    & $c$     & $d$      \\\hline
&&&&&&\\
${}^{18}$O/${}^{16}$O & quartz - forsterite     & 0-400  &  0.39600 & 3.16321 & -0.14485 & 0.003266 \\
                      & quartz - forsterite 	&400-1500&  0.00109 & 3.48189 & -0.23010 & 0.011498 \\ 
                      & quartz - enstatite 	& 0-400  &  0.31576 & 2.13612 & -0.10597 & 0.002457 \\
		      & quartz - enstatite 	&400-1500&  0.00090 & 2.39124 & -0.17466 & 0.009142 \\
                      & quartz - lizardite 	& 0-400  & -0.9984  & 2.7228  & -0.07952 & 0.001235 \\
		      & quartz - lizardite 	&400-1500& -0.08182 & 1.5739  &  0.43048 &-0.077999 \\
${}^{30}$Si/${}^{28}$Si &quartz - kaolinite\tnote {$\dagger$}     & 0-400  &  0.02003 & 0.17963 & -0.00445 & 0.000099 \\
		   	&quartz - kaolinite\tnote {$\dagger$}	&400-1500& -0.00167 & 0.19761 & -0.00941 & 0.000592 \\
		      & quartz - forsterite 	& 0-400  &  0.05381 & 0.38927 & -0.01059 & 0.000257 \\
                      & quartz - forsterite 	&400-1500&  0.00001 & 0.43433 & -0.02326 & 0.001540 \\
                      & quartz - enstatite 	& 0-400  &  0.04043 & 0.46292 & -0.01271 & 0.000258 \\
                      & quartz - enstatite 	&400-1500& -0.00017 & 0.49721 & -0.02261 & 0.001320 \\
                      & quartz - lizardite 	& 0-400  &  0.05691 & 0.62718 & -0.01970 & 0.000416 \\
                      & quartz - lizardite 	&400-1500& -0.00008 & 0.67372 & -0.03229 & 0.001635 \\
\hline
\end{tabular}
	\begin{tablenotes}
	\item [$\dagger$] \citemt{Meheut2007}.
	\end{tablenotes}
\end{threeparttable}
\end{center}
\end{table*}

Logarithmic fractionation laws deduced from the fits of the $\beta$-factors are reported in Table \ref{tab:fits}.

\subsubsection{Oxygen Fractionation.}

\begin{figure}[!hbp]
        {\includegraphics[width=120mm]{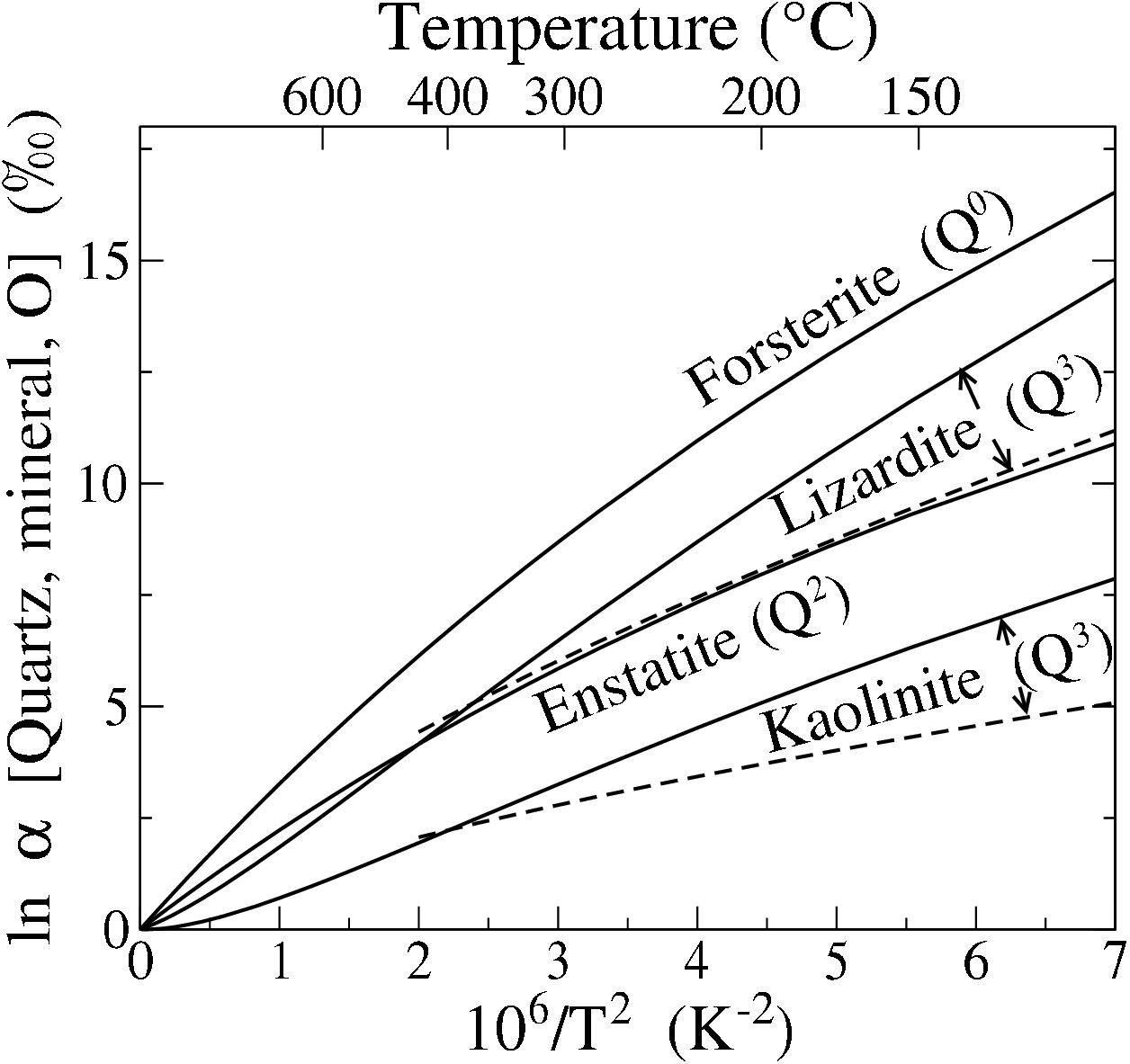}}
        \caption{Theoretical oxygen isotope fractionation 
	factors between quartz and other minerals (solid lines).
	Dashed lines: calculations of \citemt{Zheng1993} for lizardite and kaolinite. 
	In parenthesis: polymerization degree of each material.}
        \label{fig:Oqtz-min}
        \end{figure}

	The resulting quartz-mineral fractionation factors relative to ${}^{18}$O have been plotted 
in Figure \ref{fig:Oqtz-min}, for temperatures above 100$^\circ$C. 

	Before to compare this calculation to the available fractionation data, 
we can make three remarks. First, the two hydroxyl-bearing (lizardite and kaolinite) minerals 
exhibit a fractionation curve with an inflexion point at around 700$^\circ$C, on the contrary to 
forsterite and enstatite, which leads the lizardite and enstatite curves to cross each other at around 400$^\circ$C.
Indeed, because of the high vibrational frequencies of the hydroxyl units, the temperature dependence 
of the fractionations is expected to be more complicated than in anhydrous silicates (\citemp{Bottinga1973}).
 This effect is nevertheless 
limited, and does not significantly affect the linearity of the fractionation law at high temperatures. 

	Second, several studies have remarked that the oxygen isotopic content could be related to its 
silicon and cationic content. In a theoretical point of view, these empirical relations 
were justified in terms of the relative proportions of Si-O-Si, Si-O-Al and Si-O-M (M=Fe,Mg) bond-types 
in their structures, the structures having the biggest proportion of strong bonds 
being also the ones with the higher ${}^{18}$O content (by order of decreasing strength, we have 
Si-O-Si $>$ Si-O-Al $>$ Si-O-M). 
For example, \citemt{Garlick1966} gave a simple empirical relationship, 

$$\delta {}^{18}\textrm{O} = KI +C, $$
where $K$ and $C$ are parameters depending on temperature and on the isotopic composition of a reference mineral, and
$$I = \frac{\textrm{Si} + 0.58\textrm{Al equivalents}}{ \textrm{Total equivalents}},$$
where Si, Mg and Al equivalents are respectively the number of $\frac{1}{2}$(SiO$_2$), $\frac{1}{3}$(Al$_2$O$_3$), 
(MgO) units in one structure.

 For the case of the MgO-SiO$_2$ system, the order of decreasing ${}^{18}$O content should thus 
correspond to the order of decreasing Si content, and
we should have, by order of decreasing ${}^{18}$O content: quartz $>$ enstatite $>$ lizardite $>$ forsterite.
In terms of quartz-mineral fractionation, this would give: 
quartz-enstatite $<$ quartz-lizardite $<$ quartz-forsterite.
This order corresponds only partly to the order of decreasing polymerization degree, which 
 would give quartz-lizardite $<$ quartz-enstatite $<$ quartz-forsterite. 
Actually, below 400$^\circ$C, Fig.~\ref{fig:Oqtz-min} shows that
 the order given by Garlick's empirical relationship is observed, while
above this temperature, we observe the order of decreasing polymerization degree. 
Therefore, both relationships are quite approximate in the case of oxygen, 
but nevertheless correspond more or less to what is observed.

	Third, concerning the kaolinite-quartz equilibrium, the empirical relationship of \citemt{Garlick1966}
predicts that it should be 2.5 times smaller than quartz-lizardite, which is roughly what is observed. 
Kaolinite and quartz are very closely related structures. Their only difference is their octahedric layers: 
they consist of AlO$_6$ units in kaolinite, and MgO$_6$ in the case of lizardite. 
The difference between quartz-kaolinite and quartz-lizardite fractionation factors 
can thus be seen as the 
effect of a "Mg-Al substitution". We note that more recent calculation by 
\citemt{Zheng1993}, based on modified increment methods, 
which however gives quantitatively different results, also correctly 
takes into account the effect of the Mg-Al substitution (see Fig.~\ref{fig:Oqtz-min}).

% a reprendre sans doutes

\paragraph{Quartz-forsterite fractionation}

  	\begin{figure}[!hbp]
        {\includegraphics[width=120mm]{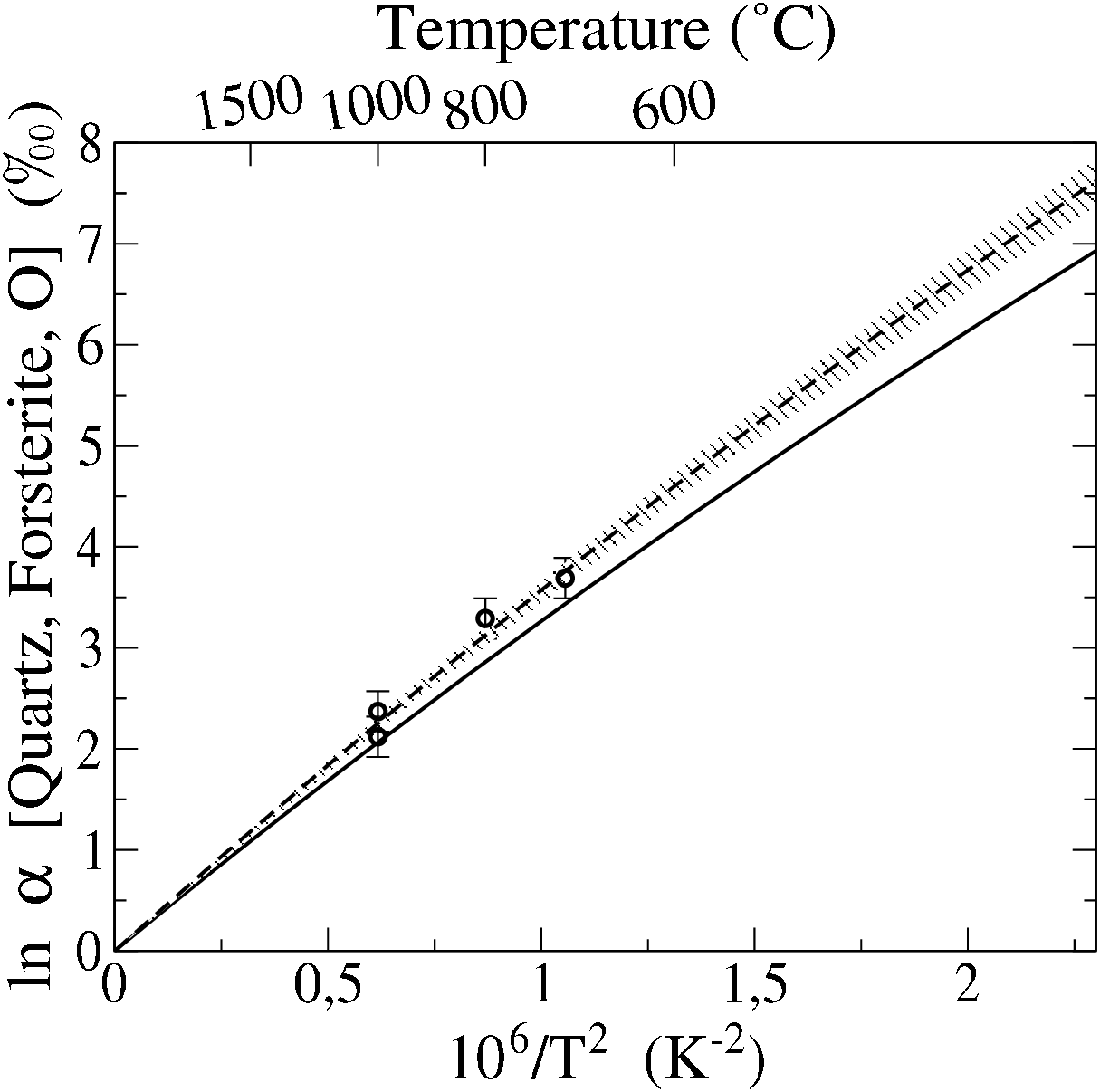}}
        \caption{Theoretical oxygen isotope fractionation factor
        between quartz and forsterite (solid line).
	The opened circles are obtained by combining forsterite/calcite data of \citemt{Chiba1989}, and  
	quartz/calcite data of \citemt{Clayton1989}. 
	Thick dashed line: semi-empirical curve proposed by \citemt{Clayton1991}; 
	Diagonal lines: corresponding error envelope.} 
        \label{fig:Oqtz-for}
        \end{figure}

The calculated quartz-forsterite equilibrium oxygen isotopic fractionation is shown Figure \ref{fig:Oqtz-for}. 
For comparison, we have reported quartz-forsterite fractionation factors obtained by combining 
forsterite/calcite data of \citemt{Chiba1989} and quartz/calcite data of \citemt{Clayton1989} 
for temperatures common to the two experiments (at  700, 800 and 1000$^\circ$C).
 The semi-empirical curve 
of \citemt{Clayton1991}, $\Delta \textrm{Qtz-Fo} = 3.790x-0.228x^2+0.0091x^3$, 
with $x=\frac{10^6}{T^2}$, is also shown for comparison, with the corresponding error envelope. 
Uncertainties on the obtained quartz-forsterite measurements are estimated from the 
analytical uncertainty given by \citemt{Chiba1989}, considered as a standard deviation. 
 
The agreement between this calculation and experiment is compatible with the 5\% error estimated 
for this kind of calculation (\citemp{Meheut2007}).

\paragraph{Enstatite-forsterite}

	\begin{figure}[!hbp]
        {\includegraphics[width=120mm]{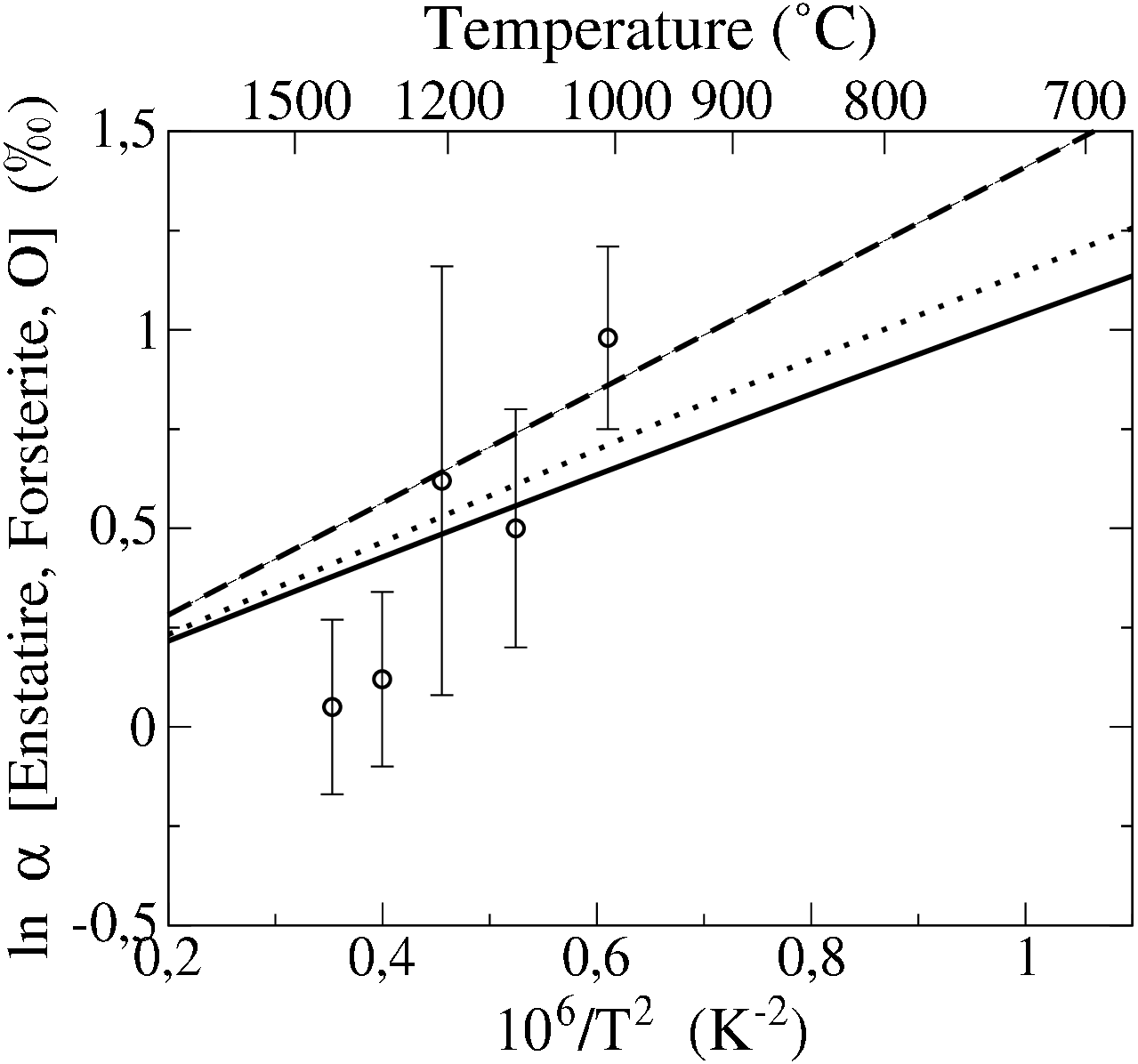}}
        \caption{Theoretical oxygen isotope fractionation factor
        between enstatite and forsterite (solid line).
	Opened circles : combination of enstatite/witherite and forsterite/witherite data from \citemt{Rosenbaum1994}
        Thick dashed lines: experimental calibration proposed by \citemt{Rosenbaum1994}.
	Pointed lines: calculations by \citemt{Kieffer1982} }
        \label{fig:OEns-for}
        \end{figure}

The calculated enstatite-forsterite equilibrium oxygen isotopic fractionation, shown Figure \ref{fig:OEns-for} can 
be compared to the experimental measurements by \citemt{Rosenbaum1994} (combining enstatite-witherite 
and forsterite-witherite experiments), and calculations by \citemt{Kieffer1982} on  
clinoenstatite-quartz are also shown for comparison. Due to the addition of the uncertainties of 
the two experiments, and to the small amplitude of the fractionation between these two minerals, 
the relative uncertainties on the experimental law is important (see \citemp{Rosenbaum1994}).
Considering that the relative error of our calculation 
is thought to be similar in forsterite-enstatite and in quartz-forsterite, due to the 
consistency of the calculation, 
we remark that the enstatite-forsterite calculation of \citemt{Kieffer1982} gives a result in better agreement 
to our calculation than the fit given by \citemt{Rosenbaum1994}. But, most importantly, given the high uncertainty on 
those different results, this independent calculation diminishes the error envelope on this fractionation law. 

% il serait preferable de proposer un rescaling

\paragraph{Quartz-lizardite fractionation}

          \begin{figure}[!hbp]
        {\includegraphics[width=120mm]{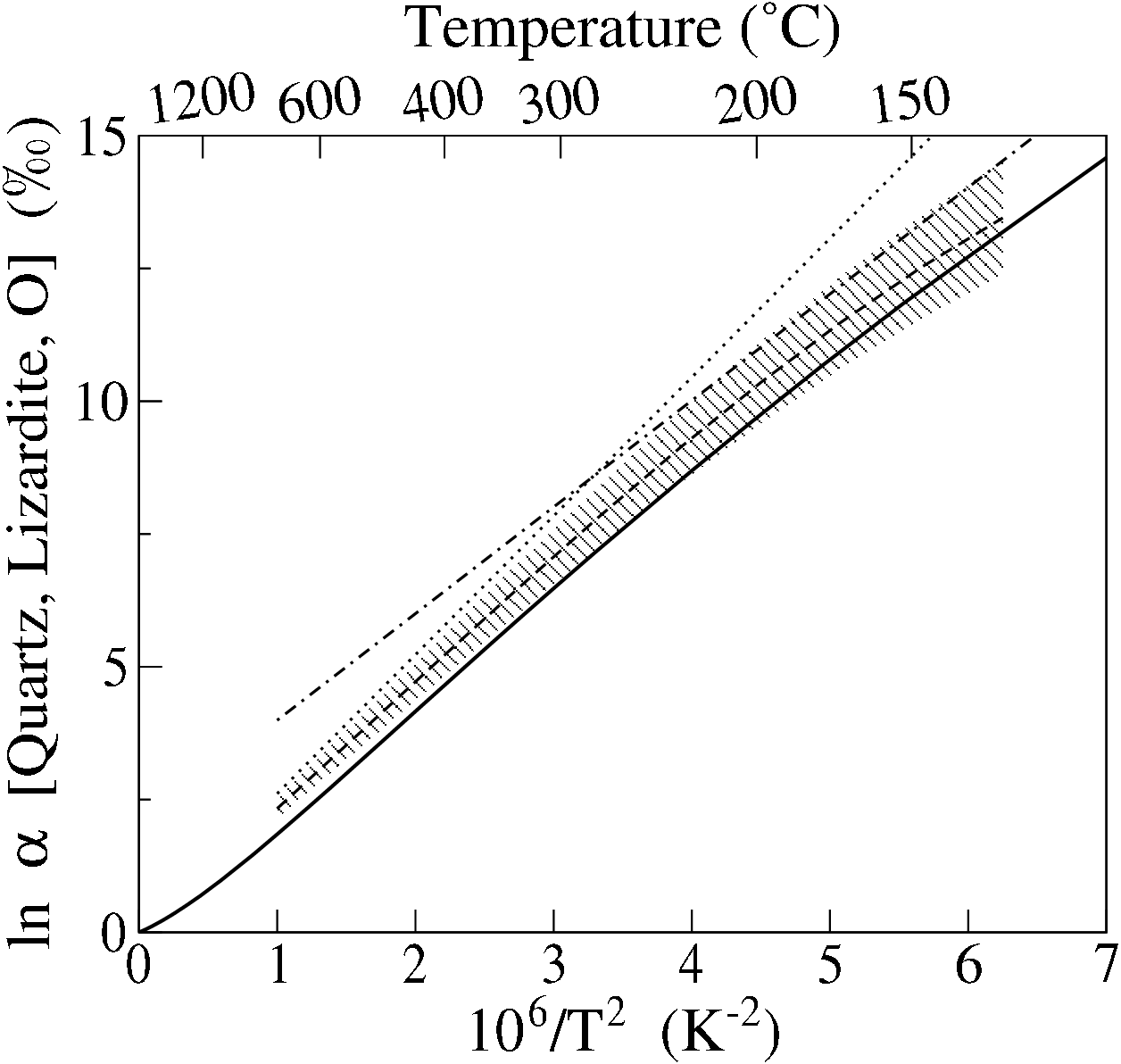}}
        \caption{Theoretical oxygen isotope fractionation factor
        between quartz and lizardite (solid line), 
        and comparison with laws obtained from different quartz-muscovite laws, following \citemt{Wenner1971} 
	(see text). Point-dotted line: law proposed by \citemt{Wenner1971}. Thick dotted line: 
	the law that we obtained from the quartz-muscovite law of \citemt{Matthews1984}. 
	Dashed line: the law that we obtained from the quartz-muscovite law of \citemt{Chacko1996},
	 and corresponding error envelope (diagonal lines).
	} 
        \label{fig:OQtz-LizHT}
        \end{figure}

The calculated quartz-lizardite oxygen isotopic fractionation, shown Figure \ref{fig:OQtz-LizHT}, 
can neither be compared to quartz-lizardite measurements nor be obtained from a combination 
of solid-solid measurements (as done in the previous cases).
Nevertheless, \citemp{Wenner1971} proposed a methodology to deduce quartz-serpentine 
fractionation laws from quartz-muscovite laws. This method is based on two assumptions. 
First, by comparing quartz/mineral fractionations in rocks containing coexisting quartz, 
muscovite and chlorite they obtain the empirical relation 
$\frac{\Delta quartz-chlorite}{\Delta quartz-muscovite}= 1.6857$. 
Second, they assume that  the fractionation factor chlorite-lizardite should be equal to zero, 
on the basis of bond-type considerations. This last hypothesis was further 
confirmed by the more precise calculations of \citemt{Zheng1993}. 
On Figure \ref{fig:OQtz-LizHT}, we show the quartz-lizardite law proposed 
on this basis by \citemt{Wenner1971}.
Following the same approach, we calculated the quartz-lizardite laws from
different sources of quartz-muscovite fractionation measurements.
A calibration of the quartz-muscovite geothermometer is proposed by
\citemt{Matthews1984} based on isotopic data of natural samples.
The corresponding empirical quartz-lizardite law that we obtained is
reported in Figure \ref{fig:OQtz-LizHT}.
More recently, \citemt{Chacko1996} realized muscovite-calcite exchange
experiments.  They adopted the combined experimental/theoretical
approach of \citemt{Clayton1991} to propose a muscovite-calcite law.
The quartz-calcite law of \citemt{Clayton1991} can be used to obtain
the quartz-muscovite law and, thus, the corresponding empirical quartz-lizardite
law we report in Figure \ref{fig:OQtz-LizHT}.
This kind of approach gives a law valid \emph{a priori} for T $>$ 400K
(\citemp{Clayton1991}).
The errors associated with the dominant coefficient of
quartz-muscovite fractionation equation being on the order of
$\pm0.10$ (1$\sigma$) (\citemp{Clayton1991}), the corresponding error
for quartz-lizardite is $\pm0.17$, due to the multiplicative
factor. The error envelope corresponding to this uncertainty is shown
on Figure \ref{fig:OQtz-LizHT}.
Our first-principles calculation is in very good
agreement with the quartz/lizardite law obtained with the \emph{a
priori} most accurate quartz-muscovite data, thus confirming the
relevance of this empirical approach.

\subsection{Silicon fractionation}

	\begin{figure}[!hbt]
        {\includegraphics[width=120mm]{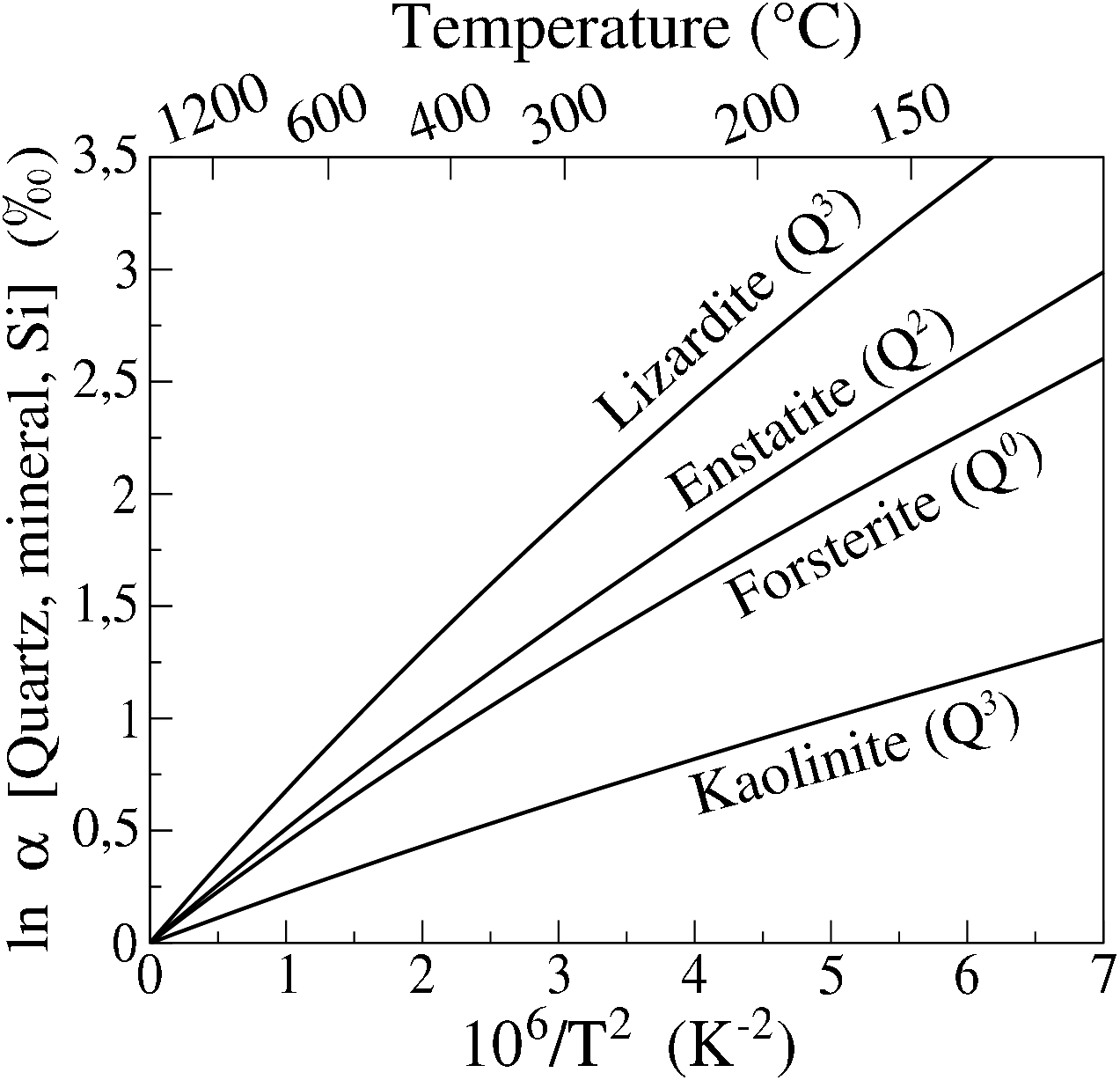}}
        \caption{Theoretical silicon isotope fractionation
        factors between quartz and other minerals (solid lines). In parenthesis: 
	polymerization degree of each material.}
        \label{fig:Siqtz-min}
        \end{figure}

        The calculated silicon fractionation curves are shown in figure \ref{fig:Siqtz-min}, for temperatures 
above 100$^\circ$C corresponding to high temperature processes.
The curves show, as expected, a more linear behavior than oxygen fractionation laws, 
in the case of kaolinite and lizardite. The fractionations can reach important values, up to 
typically 3$\permil$ around 200$^\circ$C. The kaolinite is the material with the smallest 
fractionation factor with respect to quartz. 
It is nevertheless the most interesting for comparison with low temperature data 
(see below). Comparison with experimental data, as well as with oxygen fractionation and relationship 
with polymerization degree are 
the subjects of the following sections.

\subsubsection{Low Temperature fractionation between dissolved silicon and kaolinite}

   	\begin{figure}[!hbp]
        {\includegraphics[width=120mm]{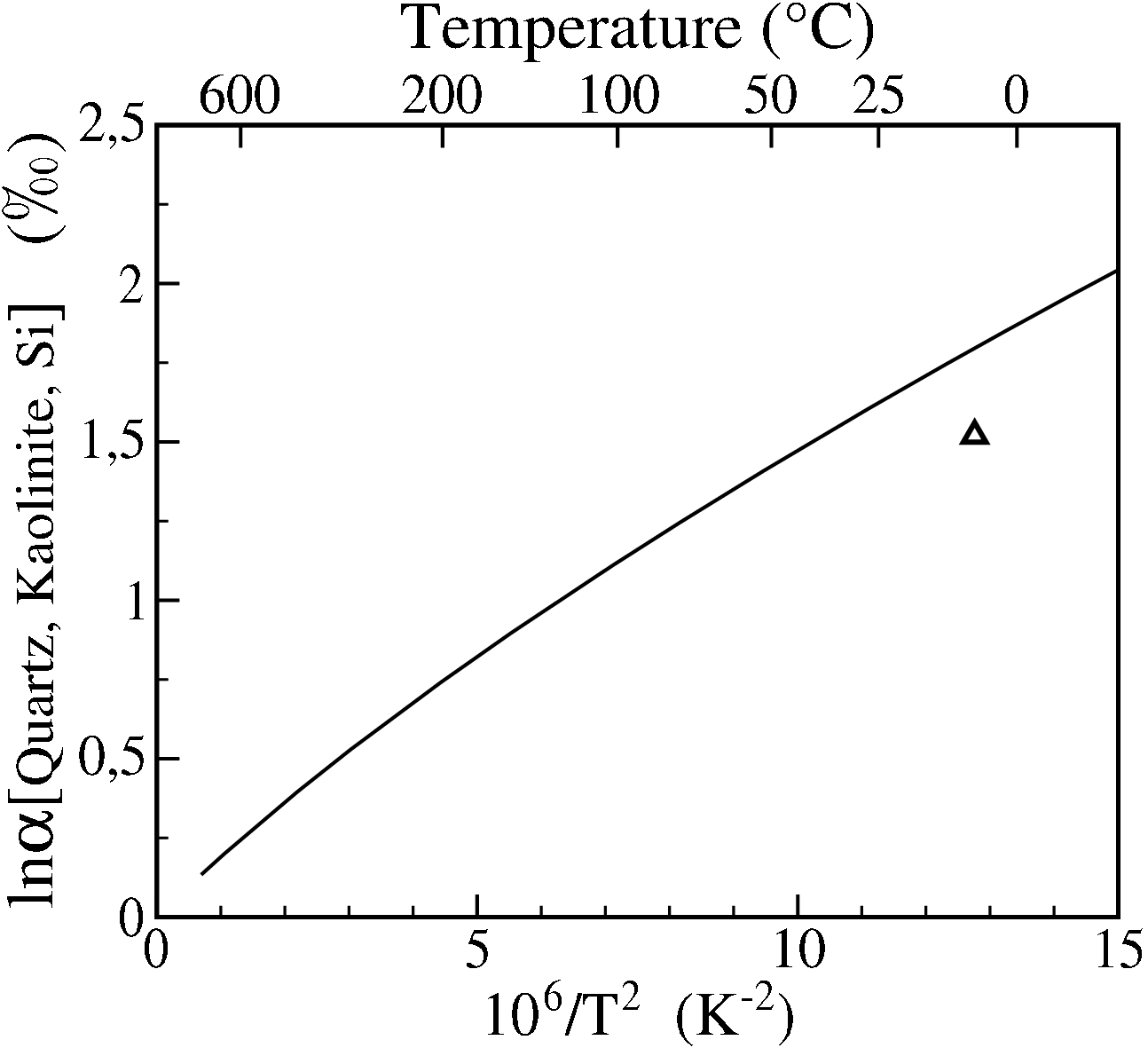}}
        \caption{Comparison of calculated quartz-kaolinite silicon fractionation with dissolved silicon-clays 
	fractionation factor measured by \citemt{Georg2007}.}
        \label{fig:kaolqtzsi}
        \end{figure}

In Figure \ref{fig:kaolqtzsi}, we report our calculated quartz-kaolinite fractionation factor.
According to \citemt{Douthitt1982},
the value of the fractionation factor between quartz and dissolved silicon 
is 0$\pm$0.2$\permil$ for temperatures going from 50 to 250$^\circ$C.
We consider that this value can be extrapolated to any temperatures, particularly lower ones,
and we compare our calculation with the fractionation between dissolved-silicon and clays
reported by \citemt{Georg2007}.
They report a value of $1.5\permil$ corresponding to 
7$\pm4^\circ$C (2$\sigma$), which is the average
temperature of the data reported.
At this temperature, our calculated fractionation is 1.8$\permil$. 
The uncertainty concerning the quartz-dissolved silicon, as well as the one potentially due to our extrapolation 
seems sufficient to explain the biggest part of the difference between the two values.
Thus, the agreement between the two values suggests that the fractionation
measured by \citemt{Georg2007} corresponds to a situation of equilibrium.
Our calculated fractionation goes from 1.87 at 0$^\circ$C to 1.58 at 30$^\circ$C, 
giving a variation of 0.3$\permil$. We remark that such a temperature dependence should be 
measurable, given the degree of analytic precision recently attained (\citemp{Georg2007}).
 
% chercher references diffusivite Si/O dans quartz

\subsubsection{High temperature data. }

        At high temperature (1400K will be taken as a reference, see \citemp{Douthitt1982}),
the calculated fractionations of silicon 
(respectively 0.11, 0.23, 0.26, 0.36 $\permil$ between quartz and kaolinite, forsterite,
enstatite, lizardite, see figure \ref{fig:Siqtz-min}), are consistent
with the fractionations of Table \ref{tab:ding}.
Our quartz-enstatite
calculation can be more precisely compared with the measurements on Lunar Rocks by \citemt{Taylor1973}.
(Table \ref{tab:taylor}). In fact, this author gives an estimate of the equilibrium temperature
(based on the oxygen fractionation)  for the system cristobalite-pyroxene, system which is similar
to the quartz-enstatite one.
This similarity is confirmed by the good agreement 
between calculated and measured oxygen fractionation factors (see Table \ref{tab:taylor})
and allows the comparison of the silicon fractionation in the two systems.
The agreement is poorer than for oxygen, our calculation underestimating by 50\% 
the measured value. 
Analytical and calculation uncertainties likely explain this discrepancy on high-T
fractionation factors. Albeit not accounting for the theoretical underestimation 
of Si fractionation factors, it is noteworthy that the temperature recorded by 
the Si and O isotopes may differ because of the different diffusivity of the 
two elements (see e.g. \citemp{Bejina1995}; \citemp{Roma2001}). 
In that view, the slowly diffusing Si isotopes could record equilibration at a higher
temperature than O isotopes.

\begin{center}
\begin{table*}
%\scriptsize
\caption{Data on silicon equilibrium fractionation from \citemt{Ding1996}, shown as a
 function of the polymerization degree of materials}
\label{tab:ding}
\begin{threeparttable}
\begin{tabular}{c|cccc|l}
\hline \hline
Type of source & \multicolumn{4}{c|}{Materials by decreasing order of polymerization} & Fractionation factors\\
                       & tectosilicate & phyllosilicate &inosilicate&nesosilicate&    \\\hline
biotite granite \tnote {$\dagger$}& quartz & biotite    &           &            & 0.2-0.4 $\permil$  \\
 granite \tnote {$\dagger$}       & quartz & muscovite  &           &            & 0.3 $\permil$ \\
biotite granite \tnote {$\dagger$}&feldspar& biotite    &           &            & 0.2$\permil$ \\
Garnet-biotite schists \tnote {$\dagger$}& & biotite    &           & garnet     & 0.6 $\permil$\\
 ore deposit \tnote {$\dagger$}   &albite &             & riebekite &            & 0.4$\permil$ \\
 ore deposit \tnote {$\dagger$}   &albite &             & aegirine  &            & 0.9$\permil$\\
this study              & quartz &kaolinite,serpentine&clino-enstatite&forsterite& \\\hline
\end{tabular}
  \begin{tablenotes}
   \item [$\dagger$] \citemt{Ding1996}
  \end{tablenotes}
\end{threeparttable}
\end{table*}
\end{center}

\begin{table*}
\begin{center}
\caption{Data on silicon equilibrium fractionation from \citemt{Taylor1973}. }
\label{tab:taylor}
\begin{threeparttable}
\begin{tabular}{c|cccc}
\hline \hline
Type of source          & equilibrium             &$\alpha({}^{30}Si)$&$\alpha({}^{18}O)$&T ($^\circ$C)\\ \hline
Lunar rocks \tnote {$\star$}&plagioclase/pyroxene  & 0.2-0.3$\permil$ & 0.3$\permil$ & 1200 \tnote{$\dagger$}\\
Lunar rocks \tnote {$\star$}&cristobalite/pyroxene & 0.52$\permil$    & 1.2$\permil$ & 1200 \tnote{$\dagger$}\\
   This study            &quartz/clinoenstatite& 0.26$\permil$    & 1.14$\permil$& 1200 \\ \hline
\end{tabular}
  \begin{tablenotes}
   \item [$\star$] \citemt{Taylor1973}
   \item [$\dagger$] estimated from  $\alpha({}^{18}O)$
  \end{tablenotes}
\end{threeparttable}
\end{center}
\end{table*}

\subsubsection{Silicon fractionation and polymerization degree.}

	According to \citemt{Grant1954}, the ${}^{30}$Si content is expected to increase with the degree of
polymerization. This can be understood on the basis of bond-type considerations. In fact, contrary to oxygen,
Si forms only O-Si bonds and, thus, the fractionation properties of silicon 
should be related only to the strength of the O-Si bonds. Since the
Si-O stretching frequencies seem to increase with the number of oxygen bonds
between SiO$_4$ tetrahedrons, the ${}^{30}$Si content should increase with the degree of polymerization.
	 On figure \ref{fig:Siqtz-min}, no clear correlation between silicon fractionation and 
polymerization degree can be seen, apart from the fact 
that all quartz-mineral laws are positive (see discussion in 3.3.1). One most singular feature is that the 
forsterite-enstatite has the sign opposite to the one expected from their polymerization degrees. 
At this point, a return on the data of Table \ref{tab:ding} seems interesting.
Apart from the biotite-garnet equilibrium, all data deal with equilibrium between tectosilicates and another silicate
of lower polymerization degree. In the framework of our system, the fact that the calculated fractionation quartz-mineral
is always positive is thus consistent with those last data. Furthermore, our calculation suggests that the 
 fractionation between phyllosilicates (Q$^\textrm{3}$) and nesosilicates (Q$^\textrm{0}$) , such as biotite/garnet data, 
may be sometimes positive 
(kaolinite-forsterite calculation), sometimes negative (lizardite-forsterite calculation). 
As well, the trend between silicon isotopic composition and silicon content observed by \citemt{Douthitt1982}
in some particular cases does not seem generalizable to any cases. 

\subsubsection{Silicon fractionation and oxygen fractionation.}

	\begin{figure}[!hbp]
        {\includegraphics[width=120mm]{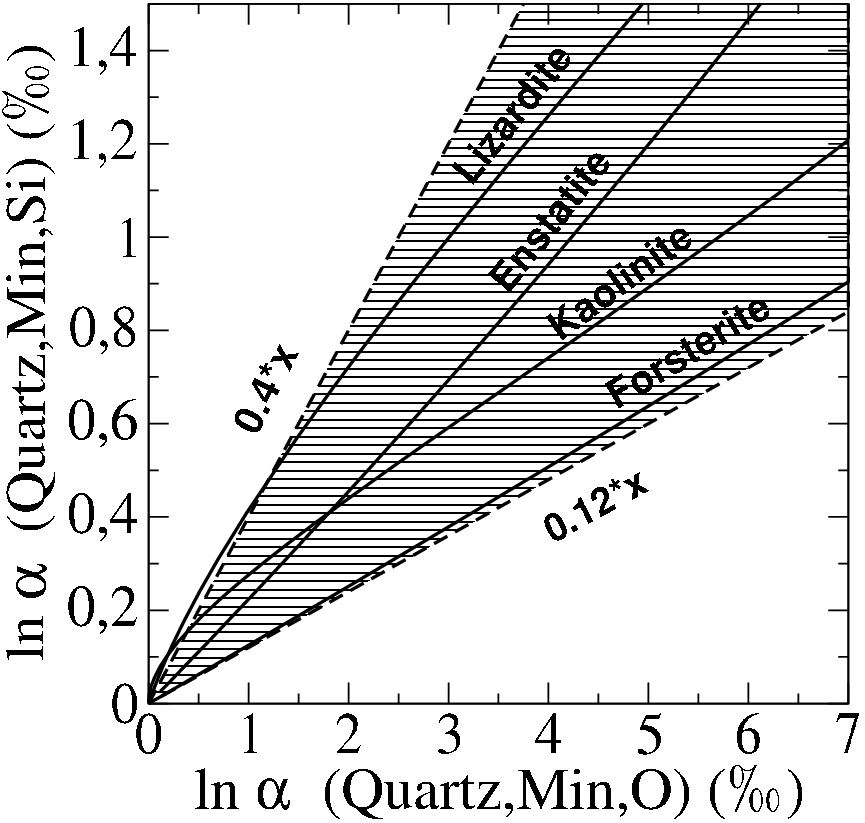}}
        \caption{Comparison between oxygen and silicon fractionation properties: quartz-mineral oxygen fractionation
	has been plotted versus quartz-mineral silicon fractionation. Horizontal lines: area between lines 
	y=0.4*x and y=0.12*x.}
        \label{fig:OSicmpHT}
        \end{figure}

To test the assumption, proposed by \citemt{Douthitt1982}, that silicon and oxygen fractionation factors may be 
roughly proportional, we have plotted on figure \ref{fig:OSicmpHT} the first as a function of the second.
More precisely, \citemt{Douthitt1982} emphasized that systematic silicon isotopic fractionations were roughly 1/3
the magnitude of concomitant oxygen isotopic fractionations at 1150$^\circ$C. In the framework of this study, 
the ratio between quartz-mineral silicon and oxygen fractionations thus goes from 0.12 (forsterite-quartz), and 
around 0.4 (lizardite-quartz equilibrium). Forsterite and lizardite are thus the two materials showing 
the most peculiar behavior from this point of view. Indeed, we have seen that 
enstatite-forsterite fractionation factors have opposite sign
for silicon and oxygen fractionation, as well as lizardite-forsterite, 
which represents the biggest failure of the assumed relationship between
fractionations of oxygen and silicon.

\subsubsection{Silicon fractionation and cationic content.}

        \begin{figure}[!hbp]
        {\includegraphics[width=120mm]{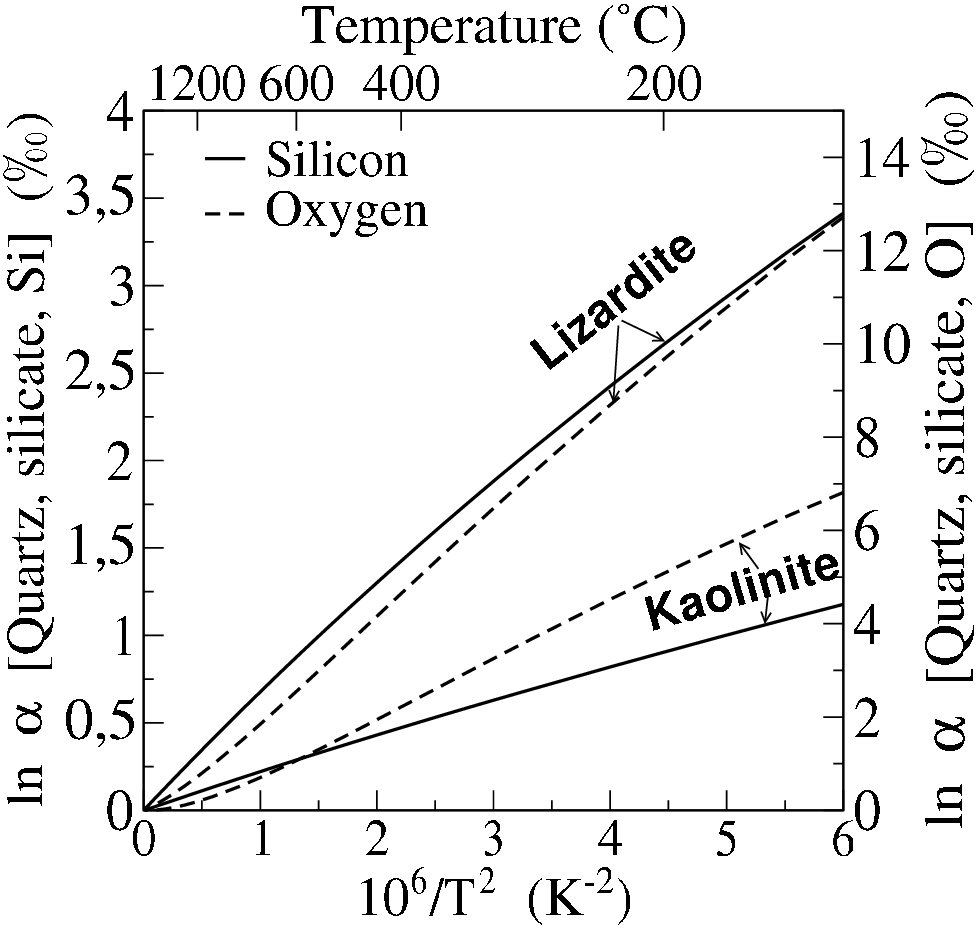}}
        \caption{Effect of the cationic content on oxygen and silicon fractionation: comparison between 
	kaolinite and lizardite silicon (solid line) and oxygen (dashed line) fractionation properties.
	Note the different scales for the two elements.}
        \label{fig:pOSiKaolLiz}
        \end{figure}

To evaluate the effect of the nature of the cations associated with Si in silicates 
 on fractionation properties, we have represented on figure 
\ref{fig:pOSiKaolLiz} both silicon and oxygen fractionations between quartz and kaolinite and lizardite. 
Kaolinite and lizardite have very similar structures. Their main difference consists in their 
octahedric layers, consisting of MgO$_6$ octahedrons in the case of lizardite,  AlO$_6$ octahedrons 
in the case of kaolinite, with three octahedrons occupied in lizardite for two in kaolinite.
Figure \ref{fig:pOSiKaolLiz} shows that the effet of cations seems similar for the isotopic fractionation
of Si and O. This suggests that oxygen fractionation, and more precisely its variations 
with cationic content, could help understand those variations in the case of silicon fractionation. 

\section{Conclusions}

	This study answers a few questions on the assertions concerning the fractionation of silicon: if the 
fractionation between quartz and other minerals is indeed always positive, a correlation between 
the degree of polymerization, or silicon content, and silicon fractionation properties is not clear. 
To conclude that tectosilicates always 
have heavier silicon content, one should make more studies on other inosilicates, like feldspars. Generally, silicon 
and oxygen fractionations are roughly correlated. The case of the enstatite-forsterite 
equilibrium  shows the limits of this law, nevertheless.
The similar effect of cation substitution for oxygen and silicon fractionations
in the system kaolinite-lizardite 
suggests that the cationic content is an important parameter for silicon fractionation properties. 

\begin{acknowledgments}

Fruitful discussions with  P. Agrinier (Universit\'e Paris VII),
F. Guyot (Universit\'e Paris VII) and F. Poitrasson (LMTG-Toulouse) are
gratefully acknowledged.
This work was supported by the French ANR project
SPIRSE. This is IPGP contribution n$^\circ$~.
Calculations were performed at IDRIS (Orsay, France), within the project n$^\circ$060411519.

\end{acknowledgments}

\clearpage

\listoffigures
\printfigures
\printtables


\clearpage

\begin{thebibliography}{47}
\expandafter\ifx\csname natexlab\endcsname\relax\def\natexlab#1{#1}\fi

\bibitem[{Balan et~al.(2001)Balan, Saitta, Mauri and Calas}]{Balan2001a}
Balan E., Saitta A.M., Mauri F. and Calas G. (2001) First-principles modeling
  of the infrared spectrum of kaolinite.
\newblock {\em Am. Mineral.\/} {\bf 86}, 1321--1330.

\bibitem[{Balan et~al.(2002)Balan, Saitta, Mauri, Lemaire and
  Guyot}]{Balan2002a}
Balan E., Saitta A.M., Mauri F., Lemaire C. and Guyot F. (2002)
  First-principles calculation of the infrared spectrum of lizardite.
\newblock {\em Am. Mineral.\/} {\bf 87}, 1286--1290.

\bibitem[{Baldereschi(1973)}]{Baldereschi1973}
Baldereschi A. (1973) Mean-Value Point in the Brillouin Zone.
\newblock {\em Phys. Rev. B\/} {\bf 7}, 5212.

\bibitem[{Baroni et~al.(2001)Baroni, de~Gironcoli and Corso}]{Baroni2001}
Baroni S., de~Gironcoli S. and Corso A.D. (2001) Phonons and related crystal
  properties from density-functional theory.
\newblock {\em Rev. Mod. Phys.\/} {\bf 73}, 515--562.

\bibitem[{Basile-Doelsch et~al.(2005)Basile-Doelsch, Meunier and
  Parron}]{Basile-Doelsch2005}
Basile-Doelsch I., Meunier J.D. and Parron C. (2005) Another continental pool
  in the terrestrial silicon cycle.
\newblock {\em Letters to Nature\/} {\bf 433}, 399--402.

\bibitem[{Bejina(1995)}]{Bejina1995}
Bejina F. (1995) {\em Measurement of atomic self-diffusion of silicon 30 in
  silicates by nuclear micro-analysis techniques. Applications to quartz and
  diopside\/}.
\newblock Ph.D. thesis, Universit\'e de Paris 11, Orsay, FRANCE.

\bibitem[{Bottinga and Javoy(1973)}]{Bottinga1973}
Bottinga Y. and Javoy M. (1973) Comments on oxygen isotope geothermometry.
\newblock {\em Earth. Planet. Sci. Lett.\/} {\bf 20}, 250--265.

\bibitem[{Chacko et~al.(1996)Chacko, Hu, Mayeda, Clayton and
  Goldsmith}]{Chacko1996}
Chacko T., Hu X., Mayeda T.K., Clayton R.N. and Goldsmith J.R. (1996) Oxygen
  isotope fractionations in muscovite, phlogopite, and rutile.
\newblock {\em Geochim. Cosmochim. Acta\/} {\bf 60}, 2595.

\bibitem[{Chiba et~al.(1989)Chiba, Chacko, Clayton and Goldsmith}]{Chiba1989}
Chiba H., Chacko T., Clayton R.N. and Goldsmith J.R. (1989) Oxygen isotope
  fractionations involving diopside, forsterite, magnetite, and calcite:
  Application to geothermometry.
\newblock {\em Geochim. Cosmochim. Acta\/} {\bf 53}, 2985.

\bibitem[{Chopelas(1991)}]{Chopelas1991}
Chopelas A. (1991) Single crystal Raman spectra of forsterite, fayalite, and
  monticellite.
\newblock {\em Am. Mineral.\/} {\bf 76}, 1100.

\bibitem[{Choudhury et~al.(1998)Choudhury, Ghose, Chowdhury, Loong and
  Chaplot}]{Choudhury1998}
Choudhury N., Ghose S., Chowdhury C.P., Loong C.K. and Chaplot S.L. (1998)
  Lattice dynamics, Raman spectroscopy, and inelastic neutron scattering of
  orthoenstatite {Mg$_2$Si$_2$O$_6$}.
\newblock {\em Phys. Rev. B\/} {\bf 58}, 756.

\bibitem[{Clayton et~al.(1989)Clayton, Goldsmith and Mayeda}]{Clayton1989}
Clayton R.N., Goldsmith J.R. and Mayeda T.K. (1989) Oxygen isotope
  fractionation in quartz, albite, anorthite and Calcite.
\newblock {\em Geochim. Cosmochim. Acta\/} {\bf 53}, 725--733.

\bibitem[{Clayton and Kieffer(1991)}]{Clayton1991}
Clayton R.N. and Kieffer S.W. (1991) {\em Stable isotope geochemistry: a
  tribute to Samuel Epstein\/}, The Geochemical Society, chapter Oxygen
  isotopic thermometer calibrations.
\newblock pp. 3--10.

\bibitem[{Clayton et~al.(1978)Clayton, Mayeda and Epstein}]{Clayton1978}
Clayton R.N., Mayeda T.K. and Epstein S. (1978) {Isotopic fractionation of
  silicon in Allende inclusions}.
\newblock In {\em Lunar and Planetary Science Conference\/}. volume~9 of {\em
  Lunar and Planetary Science Conference\/}, pp. 1267--1278.

\bibitem[{De~la Rocha et~al.(1997)De~la Rocha, Brzezinski and
  DeNiro}]{DeLaRocha1997}
De~la Rocha C.L., Brzezinski M.A. and DeNiro M.J. (1997) Fractionation of
  silicon isotopes by marine diatoms during biogenic silica formation.
\newblock {\em Geochim. Cosmochim. Acta\/} {\bf 61}, 5051--5056.

\bibitem[{De~la Rocha et~al.(2000)De~la Rocha, Brzezinski and
  DeNiro}]{DeLaRocha2000}
De~la Rocha C.L., Brzezinski M.A. and DeNiro M.J. (2000) A first look at the
  distribution of the stable isotopes of silicon in natural waters.
\newblock {\em Geochim. Cosmochim. Acta\/} {\bf 64}, 2467.

\bibitem[{Ding et~al.(1996)Ding, Jiang, Wang, Li, Song, Liu and Lao}]{Ding1996}
Ding T., Jiang S., Wang D., Li J., Song H., Liu Z. and Lao X. (1996) {\em
  Silicon Isotope Geochemistry\/}.
\newblock Geological Publishing House.

\bibitem[{Ding et~al.(2004)Ding, Wan, Wang and Zhang}]{Ding2004}
Ding T., Wan D., Wang C. and Zhang F. (2004) Silicon isotope compositions of
  dissolved silicon and suspended matter in the Yangtze River, China.
\newblock {\em Geochimica et Cosmochimica Acta\/} {\bf 68}, 205--216.

\bibitem[{Douthitt(1982)}]{Douthitt1982}
Douthitt C.B. (1982) The geochemistry of the stable isotopes of silicon.
\newblock {\em Geochim. Cosmochim. Acta\/} {\bf 46}, 1449--1458.

\bibitem[{Engelhardt et~al.(1975)Engelhardt, Zeigan, Janke, Hoebbel and
  Weiker}]{Engelhardt1975}
Engelhardt G., Zeigan D., Janke H., Hoebbel D. and Weiker W. (1975) Zur
  Abhangigkeit der Struktur der Silikatanionen in wassrigen
  Natriumsilikatlosungen vom {Na: Si verh\"altnis}.
\newblock {\em Z. anorg. allg. Chem\/} {\bf 418}, 17--28.

\bibitem[{Epstein and Taylor(1970)}]{Epstein1970}
Epstein S. and Taylor Hugh~P. J. (1970) $^{18}$O/$^{16}$O, $^{30}$Si/$^{28}$Si,
  D/H, and $^{13}$C/$^{12}$C Studies of Lunar Rocks and Minerals.
\newblock {\em Science\/} {\bf 167}, 533--535.

\bibitem[{Fujino et~al.(1981)Fujino, Sasaki, Takeuchi and
  Sadanaga}]{Fujino1981}
Fujino K., Sasaki S., Takeuchi Y. and Sadanaga R. (1981) X-ray determination of
  electron distributions in forsterite, fayalite and tephroite.
\newblock {\em Acta Cryst.\/} {\bf B37}, 513--518.

\bibitem[{Fultz et~al.(2004)Fultz, Kelley, Lee, Aivazis, Delaire and
  abernathy}]{Fultz2004}
Fultz B., Kelley T., Lee J., Aivazis M., Delaire O. and abernathy D. (2004)
  {\em Experimental inelastic neutron scattering. With reference Manual for
  DANSE - Distributer Data Analysis for Neutron Scattering Experiments\/}.
\newblock Springer-Verlag.

\bibitem[{Garlick(1966)}]{Garlick1966}
Garlick G. (1966) Oxygen isotope fractionation in igneous rocks.
\newblock {\em Earth. Planet. Sci. Lett.\/} {\bf 1}, 361.

\bibitem[{Georg et~al.(2007)Georg, Halliday, Schauble and Reynolds}]{Georg2007}
Georg R.B., Halliday A.N., Schauble E.A. and Reynolds B.C. (2007) Silicon in
  the Earth/'s core.
\newblock {\em Nature\/} {\bf 447}, 1102--1106.

\bibitem[{Grant(1954)}]{Grant1954}
Grant F. (1954) The geological significance of variations in the abundances of
  the isotopes of silicon in rocks.
\newblock {\em Geochim. Cosmochim. Acta\/} {\bf 5}, 225--242.

\bibitem[{Gregorkiewitz et~al.(1996)Gregorkiewitz, Lebech, Mellini and
  Viti}]{Gregorkiewitz1996}
Gregorkiewitz M., Lebech B., Mellini M. and Viti C. (1996) hydrogen positions
  and thermal expansion in lizardite {1-T} from Elba: a low -temperature study
  using {Rietveld} refinement of neutron diffraction data.
\newblock {\em Am. Mineral.\/} {\bf 81}, 1111--1116.

\bibitem[{Hohenberg and Kohn(1964)}]{Hohenberg1964}
Hohenberg P. and Kohn W. (1964) Inhomogeneous electron gas.
\newblock {\em Phys. Rev.\/} {\bf 136}, 864--871.

\bibitem[{Kieffer(1982)}]{Kieffer1982}
Kieffer S.W. (1982) Thermodynamics and Lattice vibrations of
  Minerals.5.Applications to phase equilibria, isotopic fractionation, and
  high-pressure thermodynamic properties.
\newblock {\em Rev. Geophys. Space Phys.\/} {\bf 20}, 827--849.

\bibitem[{Kleinman and Bylander(1982)}]{Kleinman1982}
Kleinman L. and Bylander D.M. (1982) Efficacious form for model
  pseudopotentials.
\newblock {\em Phys. Rev. Lett.\/} {\bf 48}, 1425--1428.

\bibitem[{Kohn and Sham(1965)}]{Kohn1965}
Kohn W. and Sham L. (1965) Self-Consistent Equations Including Exchange and
  Correlation Effects.
\newblock {\em Phys. Rev.\/} {\bf 140}, A1133--A1138.

\bibitem[{Matthews and Schliestedt(1984)}]{Matthews1984}
Matthews A. and Schliestedt M. (1984) Evolution of blueschists and green
  schists rocks of Sifnos, Cyclades, Greece.
\newblock {\em Contrib. Mineral. Petrol.\/} {\bf 88}, 150--163.

\bibitem[{M\'eheut et~al.(2007)M\'eheut, Lazzeri, Balan and Mauri}]{Meheut2007}
M\'eheut M., Lazzeri M., Balan E. and Mauri F. (2007) Equilibrium isotopic
  fractionation in the kaolinite, quartz, water system: predictions from
  first-principles density-functional theory.
\newblock {\em Geochim. Cosmochim. Acta\/} {\bf 71}, 3170--3181.

\bibitem[{Monkhorst and Pack(1976)}]{Monkhorst1976}
Monkhorst H.J. and Pack J.D. (1976) Special points for Brillouin-zone
  integrations.
\newblock {\em Phys. Rev. B\/} {\bf 13}, 5188--5192.

\bibitem[{Morimoto et~al.(1960)Morimoto, Appleman and Howard
  T.~Evans}]{Morimoto1960}
Morimoto N., Appleman D.E. and Howard T.~Evans J. (1960) The crystal structures
  of clinoenstatite and pigeonite.
\newblock {\em Zeitschrift f\"ur Kristallographie\/} {\bf 114}, 120--147.

\bibitem[{Noel et~al.(2006)Noel, Catti, D'Arco and Dovesi}]{Noel2006}
Noel Y., Catti M., D'Arco P. and Dovesi R. (2006) The vibrational frequencies
  of forsterite {Mg$_2$SiO$_4$} : an all-electron ab initio study with the
  crystal code.
\newblock {\em Phys. Chem. Minerals\/} {\bf 33}, 383--393.

\bibitem[{Perdew et~al.(1996)Perdew, Burke and Ernzerhof}]{Perdew1996}
Perdew J.P., Burke K. and Ernzerhof M. (1996) Generalized gradient
  approximation made simple.
\newblock {\em Phys. Rev. Lett.\/} {\bf 77}, 3865--3868.

\bibitem[{Rao et~al.(1988)Rao, Chaplot, Choudhury, Ghose, Hastings and
  Corliss}]{Rao1988}
Rao K.R., Chaplot S.L., Choudhury N., Ghose S., Hastings J.M. and Corliss L.M.
  (1988) Lattice dynamics and inelastic neutron scattering from forsterite.
  {Mg$_2$SiO$_4$} : phonon dispersion relation, density of states and specific
  heat.
\newblock {\em Phys. Chem. Minerals\/} {\bf 16}, 83--97.

\bibitem[{Reynard(1991)}]{Reynard1991}
Reynard B. (1991) Single crystal infrared reflectivity of pure {Mg$_2$SiO$_4$}
  forsterite and ({(Mg$_0.86$,Fe$_0.14$)SiO$_4$}) olivine.
\newblock {\em Phys. Chem. Minerals\/} {\bf 718}, 19.

\bibitem[{Reynold and Verhoogen(1953)}]{Reynold1953}
Reynold J. and Verhoogen J. (1953) Natural variations in the isotopic
  constitution of silicon.
\newblock {\em Geochim. Cosmochim. Acta\/} {\bf 3}, 224--234.

\bibitem[{Richet et~al.(1977)Richet, Bottinga and Javoy}]{Richet1977}
Richet P., Bottinga Y. and Javoy M. (1977) A review of hydrogen, carbon,
  nitrogen, oxygen, and chlorine stable isotope fractionation among gaseous
  molecules.
\newblock {\em Ann.~Rev.~Earth Planet.~ Sci.\/} {\bf 5}, 65--110.

\bibitem[{Roma et~al.(2001)Roma, Limoge and Baroni}]{Roma2001}
Roma G., Limoge Y. and Baroni S. (2001) Oxygen Self-Diffusion in
  $\alpha{}$-Quartz.
\newblock {\em Phys. Rev. Lett.\/} {\bf 86}, 4564--4567.

\bibitem[{Rosenbaum et~al.(1994)Rosenbaum, Kyser and Walker}]{Rosenbaum1994}
Rosenbaum J.M., Kyser T.K. and Walker D. (1994) High-Temperature oxygen isotope
  fractionation in the enstatite-olivine-{BaCO$_3$} system.
\newblock {\em Geochim. Cosmochim. Acta\/} {\bf 58}, 26.

\bibitem[{Taylor and Epstein(1973)}]{Taylor1973}
Taylor H. and Epstein S. (1973) {${}^{18}\textrm{O}/{}^{16}\textrm{O}$ and
  ${}^{30}\textrm{Si}/{}^{28}\textrm{Si}$} studies of some Appolo 15,16 and 17
  samples.
\newblock {\em Proceedings of the 4th Lunar Science Conference\/} {\bf 2},
  1657--1679.

\bibitem[{Troullier and Martins(1991)}]{Troullier1991}
Troullier N. and Martins J. (1991) Efficient pseudopotentials for plane-wave
  calculations.
\newblock {\em Phys. Rev. B\/} {\bf 43}, 1993--2006.

\bibitem[{Wenner and Taylor(1971)}]{Wenner1971}
Wenner D.B. and Taylor H.P.J. (1971) Temperatures of serpentinization of
  ultramafic rocks based on {${}^{18}$O/${}^{16}$O} fractionation between
  coexisting serpentine and magnetite.
\newblock {\em Contrib. Mineral. Petrol.\/} {\bf 32}, 165--185.

\bibitem[{Zheng(1993)}]{Zheng1993}
Zheng Y.F. (1993) Calculation of oxygen isotope fractionation in
  hydroxyl-bearing silicates.
\newblock {\em Earth. Planet. Sci. Lett.\/} {\bf 120}, 247--263.

\end{thebibliography}
\end{document}